\documentclass[conference]{IEEEtran}
\usepackage{lipsum}
\IEEEoverridecommandlockouts

\usepackage{cite}
\usepackage{amsmath,amssymb,amsfonts}
\usepackage{algorithmic}
\usepackage{graphicx}
\usepackage{textcomp}
\usepackage{xcolor}
\usepackage{braket}
\usepackage{parskip}
\usepackage[switch]{lineno}
\usepackage{nicefrac}
\usepackage{siunitx} 
\usepackage{newtxtext} % Times New Roman text
\usepackage{newtxmath} % Times New Roman math

\def\BibTeX{{\rm B\kern-.05em{\sc i\kern-.025em b}\kern-.08em
    T\kern-.1667em\lower.7ex\hbox{E}\kern-.125emX}}
\begin{document}
\setlength{\parskip}{0pt} % 1ex plus 0.5ex minus 0.2ex}
\setlength{\parindent}{0pt}

\title{CHSH Violations using Dynamic Circuits
% \\[0.1in]
% \small{\textbf{Submitted to QCE25 as Short Paper -- Experience and Application (EXAP) -- Quantum Applications Track}}
% \vspace{-0.6in}

}
\author{

 \IEEEauthorblockN{Jean-Baptiste Waring}
 \IEEEauthorblockA{\textit{\footnotesize Dept. of Electrical \& Computer Engineering} \\
 \textit{Concordia University}\\
 Montréal, Québec \\
 j\_warin@live.concordia.ca}
 \and
 \IEEEauthorblockN{Christophe Pere}
 \IEEEauthorblockA{\textit{\footnotesize Dept. of Computer Science \& Software Engineering} \\
 \textit{INTRIQ, Laval University,}\\
 Québec, Québec \\
 christophe.pere.1@ulaval.ca}
 \and
 \IEEEauthorblockN{Sébastien Le Beux}
 \IEEEauthorblockA{\textit{\footnotesize Dept. of Electrical \& Computer Engineering} \\
 \textit{Concordia University}\\
 Montréal, Québec \\
 sebastien.lebeux@concordia.ca}
\vspace{-1cm}
}

\maketitle

\begin{abstract}

Scalable quantum computing relies on high-quality, long-range entanglement, a challenge on noisy, near-term devices.
The need for practical insights for near-term algorithm design calls for trade-offs exploration in implementing dynamic circuits on current hardware. In this work, we experimentally compare three CNOT implementations for generating Bell states across varying qubit separations on a 127-qubit IBM Quantum Eagle processor (\texttt{ibm\_quebec}): a unitary (SWAP-based) approach, a dynamic approach with mid-circuit measurements and classical feedforward, and a post-processed approach. 
We use Clauser-Horne-Shimony-Holt (CHSH) inequality violations to quantify entanglement quality. We observe that, beyond 10 qubits, dynamic circuits lead to higher $|S|$ values than the unitary approach, demonstrating improved distance-dependent entanglement preservation. The post-processed approach yields the highest CHSH values, reaching $|S| > 2$ up to 13 qubits. Our results underscore the critical need for faster classical feedforward and higher readout fidelity.

\end{abstract}


\section{Introduction}

Realizing the full potential of quantum computing technologies lies in the quality of long-range entanglement, which is at the heart of quantum algorithms \cite{hoyer_quantum_2005, pham_2d_2013}. Entanglement quality typically degrades with increasing qubit separation due to noise and decoherence \cite{burnett_decoherence_2019}. This challenge is particularly pronounced in current Noisy Intermediate-Scale Quantum (NISQ) devices \cite{preskill_quantum_2018}, which often feature limited qubit connectivity and are typically restricted to nearest-neighbor interactions.

% LAQCC 
To address the distance-dependent degradation of entanglement, Local Alternating Quantum-Classical Computation (LAQCC) \cite{buhrman_state_2024}, also known as dynamic circuits \cite{corcoles_exploiting_2021}, has emerged as a promising approach to mitigate this distance-dependent degradation of entanglement.
This paradigm offers a potential pathway to mitigate the impact of connectivity limitations in current devices. LAQCC formalism can be used to construct equivalent circuits for certain quantum gates, such as the CNOT (derived in \cite{baumer_efficient_2024}), that circumvent the need for SWAP-mediated long-range unitary interactions. This is achieved through local quantum operations (i.e. between connected qubits) and classical communication (i.e. feedforward of measurement results to apply conditional quantum gates), leading to a reduction in circuit depth \cite{raussendorf_measurement-based_2003, jozsa_introduction_2005}. While fidelity ($>0.5$) certifies the presence of entanglement, the Clauser-Horne-Shimony-Holt (CHSH) inequality  \cite{clauser_proposed_nodate} offers a complementary benchmark by directly probing non-local correlations that defy classical explanation. Due to noise, entanglement can exist without violating the CHSH threshold ($|S|>2$).

% PRIOR RESEARCH ON PROCESS FIDELITY WITH LAQCC CNOT
Prior research has demonstrated dynamic circuits' benefits with respect to fidelity \cite{baumer_efficient_2024}. However, no study, to our knowledge, has directly investigated distance-dependent non-classical correlations on noisy quantum hardware using CHSH violations. 

% *************************
% **** Paper Structure ****
% *************************
In this paper, we present an experimental evaluation of dynamic CNOT gates for preserving long-range entanglement quality on a 127-qubit IBM Quantum Eagle processor (\texttt{ibm\_quebec}). We directly assess the impact of qubit separation on non-classical correlations, as measured by CHSH inequality violations, comparing three methods: a) Unitary b) Dynamic Circtuits and c) Post-Processing. Furthermore, we demonstrate that dynamic circuits significantly mitigate the distance-dependent degradation of entanglement. Our findings reveal that, for qubit separations up to 13 qubits, the post-processing approach maintains violations above the classical bound ($|S|>2$), compared to 7 qubits for the unitary approach and 3 qubits for the dynamic approach.

This paper is organized as follows: Section II provides background on dynamic circuits and CHSH inequalities. Section III details our experimental methods, including the hardware, qubit selection and circuit implementations. Section IV presents our experimental results, comparing the distance-dependent CHSH violations for dynamic, unitary and post-processed CNOTs. Section V discusses the practical implications, limitations, and future directions of our work, including a loophole-free experiment we plan to undertake.

\section{Background}
\label{sec:background}

This section reviews the concepts underpinning our study: Local Alternating Quantum-Classical Computation (LAQCC) principles underlying dynamic circuits, Bell tests, and the CHSH inequality.

\subsection{LAQCC Principles and Dynamic Circuits}
\label{subsec:laqcc_principles}
Local Alternating Quantum-Classical Computation (LAQCC) encompasses a class of circuits that implement non-local quantum operations using only local gates and classical communication \cite{buhrman_state_2024}. They are also known as dynamic circuits. This approach can enable effective all-to-all connectivity in near-neighbors architectures \cite{baumer_efficient_2024}, as it circumvent the need for deep unitary circuits to execute long-range gates. The necessity of distributing long-range entanglement efficiently is underscored by its requirement in various quantum algorithms \cite{hoyer_quantum_2005, pham_2d_2013}. Dynamic circuit implementations often involve preparing and measuring ancillary entangled states, leveraging principles related to quantum teleportation \cite{gottesman_demonstrating_1999}, using the classical outcomes to apply conditional corrections via feedforward \cite{baumer_measurement-based_2024, baumer_quantum_2024}. This requires hardware capable of mid-circuit measurement and conditional logic. Prior experimental work has established the advantage of dynamic CNOT gates over unitary implementations using process fidelity metrics \cite{baumer_efficient_2024} and has explored other dynamic circuit applications \cite{liao_achieving_2024}. Furthermore, similar principles enable remote operations between two quantum processors, as demonstrated by \cite{carrera_vazquez_combining_2024}.

\subsection{Conceptual Space-Time Diagram for a Bell Test}
\label{subsec:spacetime}

\begin{figure}[htbp] 
    \centering
    \includegraphics[width=0.55\linewidth]{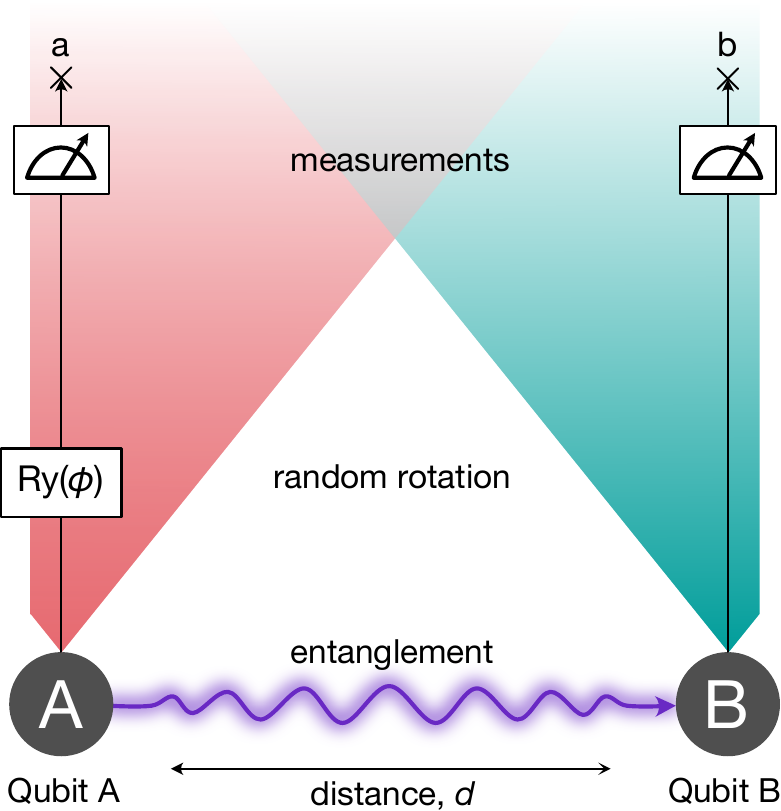} 
    \caption{Conceptual space-time diagram of a CHSH test. Entanglement is generated, a random basis is set via $R_y(\phi)$, and spacelike separated projective measurements yield outcomes \textbf{a} and \textbf{b}. Correlations between these outcomes are used to compute the CHSH parameter $S$.}
    \label{fig:spacetime_conceptual}
\end{figure}

Tests of Bell's theorem \cite{bell_einstein_1964}, motivated by the Einstein-Podolsky-Rosen paradox \cite{einstein_can_1935}, probe the conflict between quantum mechanics and local realistic theories. Figure \ref{fig:spacetime_conceptual} provides a conceptual illustration of one such test, where measurements on a spatially separated entangled pair are performed. Ideally, spacelike separation ensures that measurement settings cannot influence each other via classical communication. However, on-chip experiments typically operate well within the light cone, creating a so-called "loophole". In \cite{storz_loophole-free_2023}, a loophole-free Bell inequality violation is demonstrated using custom superconducting circuits. Instructions on how to run CHSH inequality experiments have also been made available by IBM \cite{noauthor_chsh_nodate}. However, to our knowledge, it has not been used as a way to benchmark long-range entanglement schemes.

\subsection{CHSH Inequality}
\label{subsec:chsh}
The Clauser-Horne-Shimony-Holt (CHSH) inequality \cite{clauser_proposed_nodate} offers a practical experimental test of local realism. It defines a parameter $S$ based on correlations between measurement outcomes of two parties (A, B) choosing between two settings each. Local realistic theories predict $|S| \le 2$, while quantum mechanics allows for violations up to $|S| \le 2\sqrt{2}$ for certain entangled states and measurement settings. An experimental result $|S| > 2$ thus demonstrates correlations incompatible in principle with local realism. Such violations of Bell inequalities have been demonstrated using superconducting qubits previously, such as in \cite{zhong_violating_2019}. However, these demonstrations do not rely on general-purpose, programmable, gate-based quantum superconducting computers.

\section{Method}
\label{sec:method}
In this section, we describe our experimental approach to investigate CHSH inequality violations using different CNOT implementations on gate based quantum computing hardware.

We present the quantum circuits used for performing CHSH experiments, the specific CNOT implementations under test, the hardware used and its resource selection criteria, optimizations used to improve circuit execution on the target hardware and finally, error mitigations applied to the circuit measurement outcomes.

\begin{figure*}[ht]
    \centering
    \includegraphics[width=\textwidth]{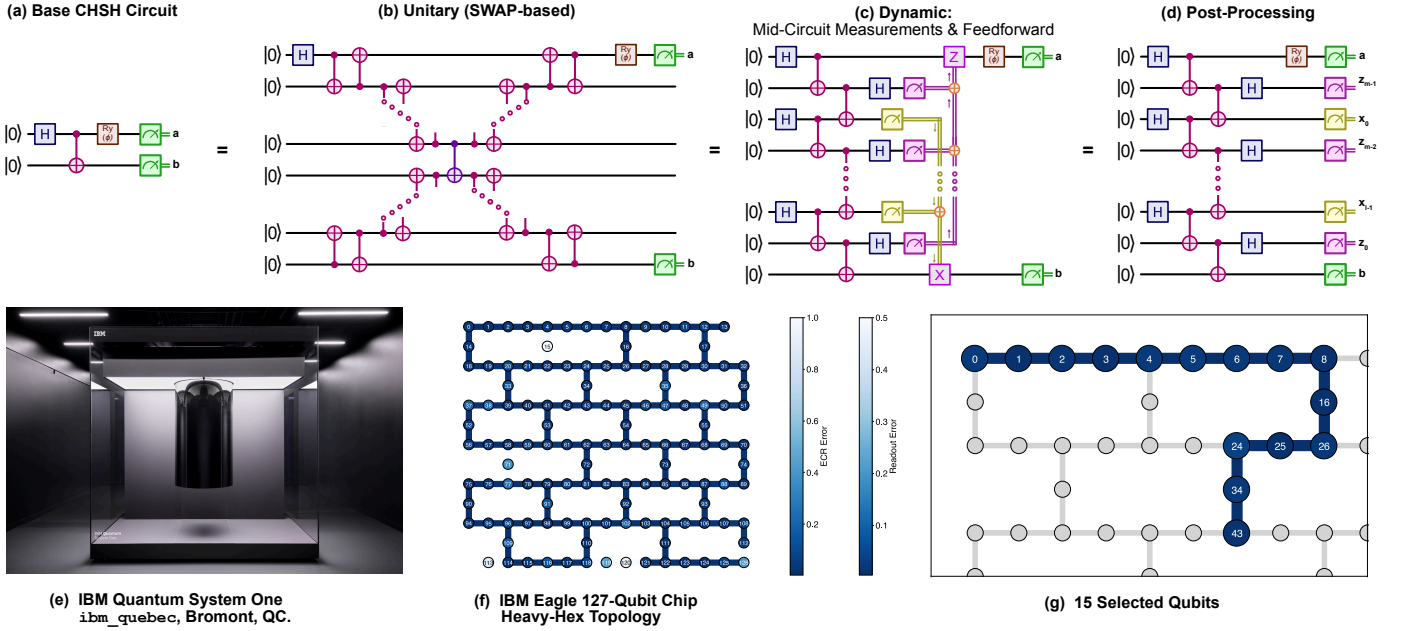}
    \caption{Circuit diagrams, hardware, and qubit selection for the CHSH experiments. (a) Base CHSH circuit: Bell state preparation using Hadamard and CNOT gates, followed by $R_y(\phi)$ rotation and measurement yielding outcomes \textbf{a} and \textbf{b}. (b) Unitary (SWAP-based) CNOT Implementation: Long-range CNOT implemented using SWAP gates (composed of CNOTs) and a central CNOT (purple). (c) Dynamic CNOT Implementation: Uses mid-circuit measurements and conditional gates ($Z$, $X$) for classical feedforward based on measurement outcomes. (d) Post-Processing Implementation: Uses the same quantum circuit structure as (c) but employs post-processing. Outcomes \textbf{a} and \textbf{b} are retained when the element-wise XOR of the $z$ register ($z_0 \oplus ... \oplus z_{m-1}$) and the $x$ register ($x_0 \oplus ... \oplus x_{l-1}$) are both $0 \equiv \texttt{FALSE}$. (e) IBM Quantum System One hosting the \texttt{ibm\_quebec} processor in Bromont, QC (image courtesy of \cite{noauthor_platform_nodate}). (f) IBM Eagle 127-Qubit Chip Heavy-Hex Topology. The color bar indicates typical ECR (two-qubit entangling gate) and readout error rates across the device. (g) The specific selection of 15 connected qubits used for the experiments, colored according to the same error scale. \textit{We thank Elisa Bäumer from IBM Zürich for inspiring the drawing of our quantum circuit diagrams.}}
    
    \label{fig:circuit_diagrams}
\end{figure*}

\subsection{CHSH Experiment}
\label{subsec:chsh_experiment}

The core experiment aims to quantify entanglement quality using a CHSH inequality \cite{clauser_proposed_nodate, noauthor_chsh_nodate}:
\begin{equation}
S = \braket{A_1 B_1} + \braket{A_1 B_2} + \braket{A_2 B_1} - \braket{A_2 B_2} \leq 2
\label{eq:chsh_general} % Added a label for potential referencing
\end{equation}
where $A_i$ and $B_i$ represent measurement operators for the first and second qubit, respectively. Experimentally, we evaluate this inequality by computing expectation values for four Pauli observables: $\braket{XX}$, $\braket{XZ}$, $\braket{ZX}$, and $\braket{ZZ}$.

To implement this on a gate-based quantum system, we focus on implementations of the circuit shown in Figure~\ref{fig:circuit_diagrams}a \cite{noauthor_chsh_nodate}. This circuit: a) prepares a Bell state $\nicefrac{1}{\sqrt{2}}(\ket{00} + \ket{11})$ between the first (control) and last (target) qubit of the selected chain using a Hadamard gate followed by a CNOT gate b) applied a rotation $R_y(\phi)$ to the control qubit, effectively changing its measurement basis relative to the target qubit c) measures both qubits in the computational basis.

To evaluate the CHSH parameter across different relative measurement bases, the phase $\phi$ is swept, and a record of each sum of observable is used to obtain a scattered estimate of $S(\phi)$, defined as:
\begin{equation}
S(\phi) = \braket{ZZ}_\phi - \braket{ZX}_\phi + \braket{XZ}_\phi + \braket{XX}_\phi
\label{eq:chsh_pauli} % Added a label
\end{equation}
This function is expected to be periodic with respect to $\phi$, and we define $\max(|S(\phi)|)$ to be the maximum value of $S(\phi)$ over the range of $\phi$ values. It is admitted that values of $|S| > 2$ indicate a violation of the CHSH inequality, which in a loophole-free experimental context, leads to the conclusion that the system exhibits correlations beyond what is explainable through local-hidden variable theories. \textbf{In the context of this paper, our goal is not to claim a demonstration of this conclusion, but rather, to use $S$ as a metric to evaluate the entanglement quality achieved by the target system whilst comparing three CNOT implementations.}

\subsection{CNOT Implementations}
\label{subsec:cnot_implementations}

We compare three distinct methods for implementing the CNOT gate between the distant control and target qubits of a CHSH circuit:
\subsubsection{Unitary (SWAP-based)}
This baseline approach, illustrated in Figure \ref{fig:circuit_diagrams}b, uses a sequence of SWAP gates to iteratively exchange the state of the control qubit with its neighbors until it is adjacent to the target qubit. The CNOT gate (purple in the figure) is then applied directly between the now-adjacent qubits. Additional SWAP gates are required to return the qubits to their original positions.

\subsubsection{Dynamic (Mid-Circuit Measurements and Feedforward)} 
This implementation, shown conceptually in Figure \ref{fig:circuit_diagrams}c, utilizes the LAQCC principle as proposed by \cite{buhrman_state_2024} and demonstrated experimentally in \cite{baumer_efficient_2024}, \cite{carrera_vazquez_combining_2024} and \cite{baumer_quantum_2024}. It involves preparing ancilla qubits in Bell pairs distributed along the chain between the control and target qubits. Mid-circuit measurements are performed on these ancillary qubits (registers $z$ and $x$). Based on the outcomes of these measurements, classical feedforward logic determines whether conditional $X$ and/or $Z$ correction gates need to be applied to the source and/or target qubits to complete the CNOT operation.

\subsubsection{Post-Processing}
This implementation, shown in Figure \ref{fig:circuit_diagrams}d, uses the same underlying quantum circuit structure as the dynamic implementation. However, conditional gates are removed, and mid-circuit measurements are delayed to occur with the final measurements of the source and target qubits. We then apply a post-processing filter to the measurement outcomes. We retain only those shots where the element-wise XOR sum of the bits in the ancillary $x$ register and the element-wise XOR sum of the bits in the ancillary $z$ register are both equal to zero ($0 \equiv \texttt{FALSE}$). This isolates the experimental outcomes corresponding to the ideal case where no active correction gates would have been necessary. This approach is used to evaluate the overhead associated with implementing mid-circuit measurements and feedforward on current hardware.

\subsection{Experimental Setup}

\subsubsection{Hardware \& Qubit Selection}
\label{subsec:hardware}

All experiments are performed on the 127-qubit IBM Quantum Eagle processor \texttt{ibm\_quebec}, a superconducting transmon device with a heavy-hexagonal topology, located in Bromont, Québec, and accessed via the IBM Quantum platform (Figure \ref{fig:circuit_diagrams}e-f). For the distance-dependent measurements, we select a linear chain of qubits (as shown in Figure \ref{fig:circuit_diagrams}g) chosen based on the device's calibration data. The experiments utilize the \texttt{Qiskit} framework \cite{treinish_qiskitqiskit-metapackage_2023}.
An initial layout specifying the exact sequence of physical qubits forming the linear chain is provided to the transpiler (Figure \ref{fig:circuit_diagrams}g). We only execute circuits for which the final physical qubit layout, determined by \texttt{circuit.layout.final\_index\_layout()}, contains precisely the same set of qubits specified in the initial layout. This ensures that all three CNOT methods are evaluated on the identical physical qubits for a given distance.

Due to the use of mid-circuit measurements and conditional logic in the dynamic implementation, the Qiskit Runtime Estimator primitive is not suitable. Instead, we use a custom workflow based on the Qiskit Runtime Sampler primitive. The Sampler collects the raw measurement outcomes (bitstrings) for all circuits. Expectation values for the required Pauli operators are then computed classically from these bitstring distributions after applying M3 readout error mitigation.

\subsubsection{Error Mitigation}
\label{subsec:mitigation}

To mitigate errors inherent to the NISQ hardware, we employ two primary techniques. First, readout assignment errors are corrected using the matrix-free measurement mitigation (M3) method provided by the \texttt{mthree} package \cite{nation_scalable_2021}. Calibration circuits are run to characterize the readout noise, and a correction matrix is applied during post-processing to the raw measurement counts. Second, to reduce decoherence effects during qubit idle times, which can be significant in deeper circuits like the unitary implementation, dynamical decoupling is applied \cite{viola_dynamical_1999, ezzell_dynamical_2023, tripathi_suppression_2022}. We use sequences of X ($\pi$-pulse) gates inserted by the Qiskit transpiler's pass manager, specifically utilizing the \texttt{PadDynamicalDecoupling} pass, configured with the hardware's timing constraints. Those mitigation techniques are applied to all three CNOT implementations using the same settings.
\section{Results}
\label{sec:results}

\begin{figure*}[!ht]
    \centering
    \includegraphics[width=1\linewidth]{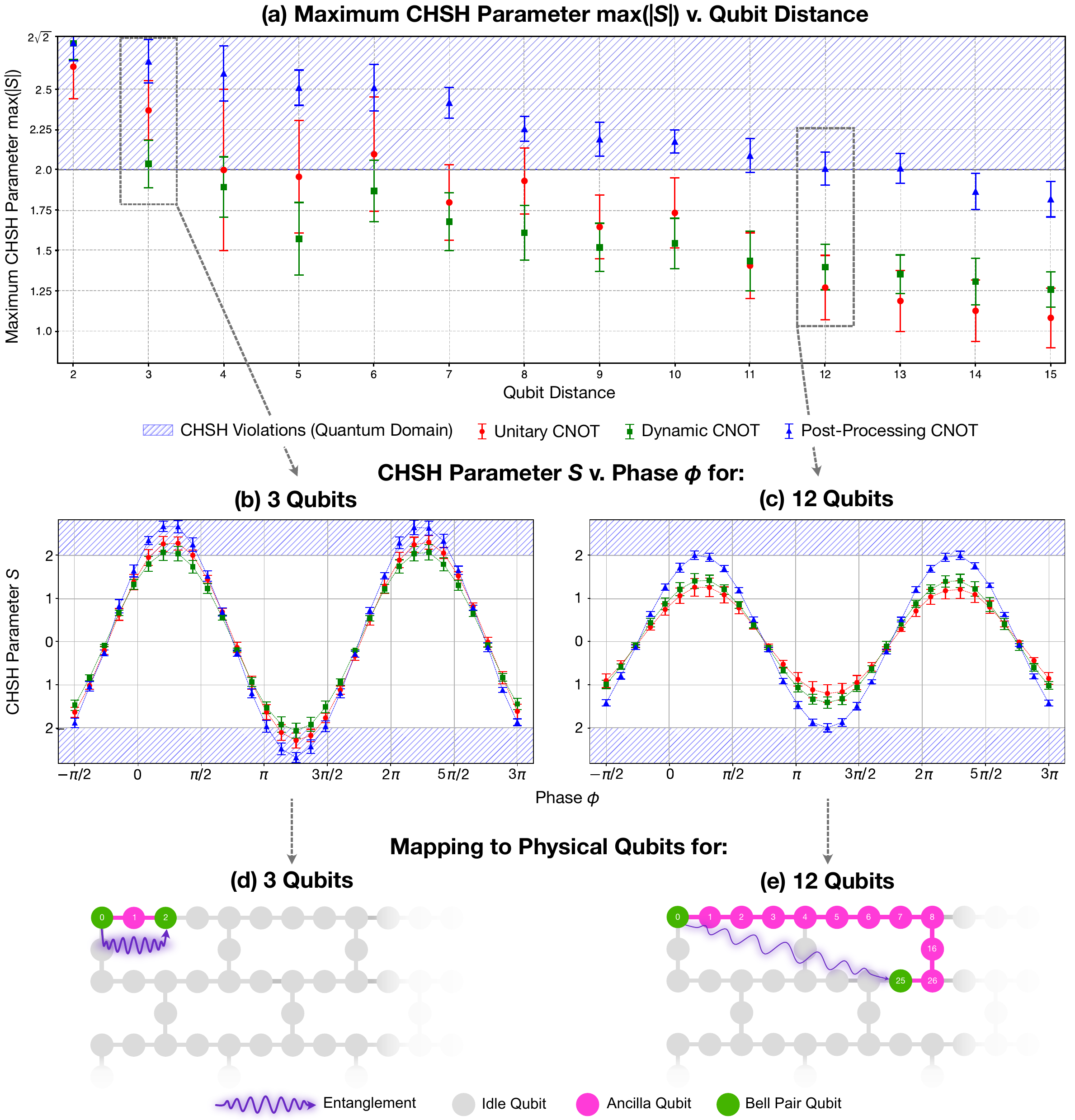}
    \caption{Experimental results of CNOT implementations for CHSH experiments across varying distances on the 127-qubit IBM Quantum Eagle processor (\texttt{ibm\_quebec}). (a) Maximum CHSH parameter, $\max(|S|)$, as a function of qubit distance (path length). Data points show the baseline unitary implementation ("Unitary CNOT", red circles), the dynamic implementation with classical feedforward ("Dynamic CNOT", green squares), and the post-processed implementation ("Post-Processing CNOT", blue triangles). Error bars indicate standard deviation. The shaded region ($|S| > 2$) denotes the Quantum Domain. (b, c) CHSH S-Curves obtained for 3-qubit and 12-qubit distances, respectively, for the three CNOT implementations as a function of the phase ($\phi$). (d, e) Qubit mappings on \texttt{ibm\_quebec}'s heavy-hexagonal processor topology for the 3-qubit and 12-qubit CHSH experiments, respectively.\\[0.5in]}
    
    \label{fig:overall_comparison}
\end{figure*}

\subsection{Experimental Setup}

To characterize CHSH violations, we sweep the phase $\phi$ through 31 points linearly spaced from $-\pi/2$ to $3\pi$. For each phase point and each required Pauli measurement basis ($XX$, $XZ$, $ZX$, $ZZ$), we execute the corresponding circuit for 10,000 shots. In summary, each experiment is comprised of $31\times4 = 124$ circuits. The expectation values $\langle XX \rangle$, $\langle XZ \rangle$, $\langle ZX \rangle$, and $\langle ZZ \rangle$ are obtained from these measurements. These expectation values are then used to calculate the a CHSH parameters, $S = \langle ZZ \rangle - \langle ZX \rangle + \langle XZ \rangle + \langle XX \rangle$. To ensure statistical robustness, the entire experiment is repeated multiple times for each condition; the results plotted in Section \ref{sec:results} are calculated from statistics over $n=20$ sampled repetitions per data point.

The post-processed experiment allows for a fairer assessment of the entanglement quality achievable by dynamic circuits by removing feedforward-related errors, it is important to note that this method is not scalable, as the fraction of retained shots decreases exponentially with the number of ancillary qubits (and thus distance). However, for the distances studied in this work (up to 15 qubits) and the number of shots used (10,000 per observable), this post-processing yields sufficient statistics for analysis.

\subsection{Overall Performance Comparison}
Figure \ref{fig:overall_comparison}a reports the maximum achieved CHSH parameter, $\max(|S|)$, as a function of qubit distance (chain length) for the three CNOT implementations. The baseline unitary implementation ("Unitary CNOT", red circles) shows high violations at short distances, with $\max(|S|) \approx 2.64$ at 2 qubits and $\max(|S|) \approx 2.37$ at 3 qubits. However, it exhibits a rapid reduction with distance, resulting in $\max(|S|)$ below the classical bound ($|S| \le 2$) for distances beyond approximately 6 qubits. This threshold ($|S| = 2$) represents the maximum correlation strength predicted by theories based on local realism; values below this bound do not rule out classical explanations for the observed measurement statistics. In contrast, the dynamic implementation with feedforward ("Dynamic CNOT", green squares) demonstrates a lower maximum CHSH value at 3 qubits ($\max(|S|) \approx 2.03$), slightly above the classical bound and significantly lower than the unitary method. A crossover point exists around 11 qubits; for longer distances, the dynamic approach results in higher $\max(|S|)$ values compared to the unitary method. The data indicate a slower rate of decrease with distance for the dynamic circuit compared to the unitary method, although its $\max(|S|)$ values are close to or below 2 for most distances shown (e.g., $\max(|S|) \approx 1.43$ at 11 qubits). The post-processed implementation ("Post-processed CNOT", blue triangles) achieves the highest $\max(|S|)$ values across all measured distances, reaching a maximum of $\max(|S|) \approx 2.67$ at 3 qubits. It displays the slowest rate of decrease with distance among the three methods and demonstrates CHSH violations ($|S| > 2$) up to 13 qubits, where the mean $\max(|S|) \approx 2.01$. This quantitative difference underscores the advantage of dynamic strategies in preserving non-classical correlations over longer distances compared to the conventional unitary method on this hardware.

\subsection{Detailed Analysis at 3 \& 12 Qubit-Distances}

Highlighting the differences at longer distances, specifically at 12 qubits, the maximum CHSH value, $\max(|S|)$, for the dynamic implementation is approximately 0.12 higher than that of the unitary approach ($\approx 1.39$ vs. $\approx 1.27$), while the post-processed value is near the classical bound at $\approx 2.01$. 

To provide a more detailed view of the CHSH violation behavior, Figures \ref{fig:overall_comparison}b and \ref{fig:overall_comparison}c plot the measured S-parameter against the measurement basis angle $\phi$ for 3-qubit and 12-qubit separations, respectively. As expected, $S$ exhibits a sinusoidal dependence on $\phi$, reflecting the changing projection onto the measurement basis controlled by the $R_y(\phi)$ gate, with extrema occurring at phases $\phi \approx (\nicefrac{4k+1}{4})\pi,\;k \in \mathbb{Z}$.

For the 3 qubit experiments, both the unitary (red circles) and post-processed (blue triangles) implementations demonstrate maximum violations well above the classical bound ($S=2$), indicative of strong non-classical correlations in principle. The dynamic implementation (green squares) has oscillations with a noticeably lower amplitude, a finding consistent with the lower $\max(|S|)$ value shown in Figure \ref{fig:overall_comparison}a; its peak value is just slightly above the classical bound.

The contrast between the implementations is more pronounced at the 12-qubit distance (Figure \ref{fig:overall_comparison}c). The unitary implementation demonstrates significant degradation; the oscillation amplitude is suppressed, and the maximum S-value is below 2, suggesting a substantial loss of the necessary non-classical correlations. The dynamic implementation, while also having a peak below the classical bound, features a visibly larger oscillation amplitude and a higher maximum S-value compared to the unitary approach at this distance. The post-processed implementation clearly shows the best performance, with the largest oscillation amplitude and a maximum S-value close to the classical bound. This visual evidence confirms its superior entanglement preservation capability over this longer distance.

\section{Discussion}
\label{sec:discussion}

\subsection{Dynamic Circuits vs. Post-processing Trade-offs}
\label{subsec:dynamic_vs_postselected}

Our experiments highlight a key trade-off in implementing dynamic circuits on current NISQ hardware. The dynamic approach, while scalable in principle, underperforms unitary circuits at shorter distances (below approx. 11 qubits on \texttt{ibm\_quebec}, see Figure \ref{fig:overall_comparison}a-b) due to overhead from mid-circuit measurements and classical feedforward (a measurement typically requires approximately 835 ns on \texttt{ibm\_quebec} compared to 595 ns for an ECR gate and 60 ns for an $X$ gate). Furthermore, classical feedforward duration is not known at transpile time and renders the use of dynamical decoupling difficult during its course. In contrast, the post-processing achieves the highest CHSH values by circumventing these errors, demonstrating dynamic circuits' potential.

\subsection{Practical Implications (EXAP)}
\label{subsec:practical_implications}
This practical experimental comparison leads us to emphasize the following implications and future directions relevant to both researchers and practitioners within the quantum computing community:
\subsubsection{Hardware Development}

Our work quantitatively underscores that the limitations of mid-circuit measurement and classical feedforward represent a major bottleneck preventing dynamic circuits from achieving their theoretical potential relative to unitary methods across all distances. We expect that a focus on faster classical feedforward, real--time application of dynamical decoupling pulses and higher-fidelity measurements will lead to significant improvements in the overall capabilities of dynamic circuits. Our method could be extended to quantify which of these improvements would be most impactful for a given architecture.

\subsubsection{Transpiler Gap}
\label{subsubsec:transpiler_gap}

Standard transpilers, including Qiskit's, primarily focus on optimizing unitary circuits, typically relying on SWAP gate insertion to resolve connectivity constraints (such as in the SABRE heuristic \cite{li_tackling_2019} or using Reinforcement Learning \cite{kremer_practical_2025}). However, they currently lack the built-in capability to automatically synthesize dynamic circuit alternatives for long-range operations or assess the associated performance trade-offs against SWAP chains based on hardware calibration.

To enable such automated, hardware-aware selection between unitary and dynamic routing, further research and development integrated into compiler workflows are needed. Directions we find particularly promising involve exploring techniques based on teleportation routing \cite{hillmich_exploiting_2021} and frameworks for automatic compilation and optimization of dynamic circuits, such as AC/DC \cite{niu_acdc_2024}.

\subsection{Proposed Loophole-Free Experiment}
\label{subsubsec:loophole_free}
Our experiments are subject to a communication (locality) loophole. The time required for applying a single entangling gate (approximately \SI{600}{\nano\second}) converts to approximately \SI{180}{lightmeters}. Consequently, the measurement settings and outcomes are not spacelike separated on a centimeter-scale experiment like ours, allowing for potential classical communication channels that could explain correlations up to the classical bound of $S=2$. To perform a more fundamental test of non-locality using entanglement generated via dynamic circuits, closing this loophole necessitates physically separating the entangled qubits over large distances. This result could be achieved by coupling two quantum processors in a similar way to \cite{carrera_vazquez_combining_2024} with a sufficiently long (dozens of km) optical link to ensure spacelike separation of the experiments. To the best of our knowledge, while prior work has demonstrated CHSH violations using superconducting qubits \cite{zhong_violating_2019, storz_loophole-free_2023}, this would be the first demonstration of a loophole-free CHSH experiment on combined, gate-based superconducting quantum processors.
\section{Conclusion}
\label{sec:conclusion}

Distributing high-quality entanglement over long distances is essential for scaling quantum computation, yet it remains challenging on current hardware. We experimentally investigate this challenge by comparing SWAP-based unitary circuits with dynamic and post-selected implementations of CNOT gates on a 127-qubit IBM quantum processor, uniquely employing CHSH inequality violations to assess distance-dependent entanglement quality. Our CHSH results show that dynamic circuits outperform their unitary counterparts in mitigating the degradation of entanglement with distance, consistent with findings from prior fidelity studies. Furthermore, by comparing dynamic circuits to a post-processed implementation, we characterize the significant performance limitations imposed by mid-circuit measurement and feedforward overhead on current hardware and show the performance gap that dynamic circuit implementations on IBM hardware need to close to be competitive with post-processing and significantly outperform their unitary counterpart.

This work demonstrates that CHSH inequality violations serve as a relevant benchmark for evaluating long-range entanglement strategies on near-term devices, complementing traditional fidelity measures by directly probing non-classical correlations. Our findings provide practical insights into the performance trade-offs, informing choices for algorithm designers while highlighting critical areas for hardware improvement—particularly the need for faster classical feedforward and higher-fidelity mid-circuit measurements.  Our results on IBM Eagle hardware suggest an advantage for dynamic circuits over their unitary counterparts. This work establishes a methodology for future benchmarking, beginning with a planned experiment on IBM's newer Heron processor to evaluate advances in dynamic circuit capabilities. Addressing these hardware limitations and bridging the gap in quantum compilers to automatically leverage dynamic circuits remain key future directions, ultimately enabling more robust long-range entanglement distribution for advanced quantum applications and potentially loophole-free tests of non-locality across networked processors.

\vspace{-0.15cm}
\section*{Acknowledgements}

We would like to thank the \textit{Plateforme d'Innovation Numérique et Quantique du Québec} (PINQ2) for the access to the machine \texttt{ibm\_quebec} and the computation time needed for this study. 

\newpage
\bibliographystyle{IEEEtran}
\bibliography{references}

\end{document}